\pdfoutput=1
\documentclass[11pt,a4paper,twoside,openright,closeany,textany]{book}
\usepackage{makeidx}
\usepackage{graphicx}
\usepackage{paralist}
\usepackage{caption} % to set figure caption size
\usepackage{amsmath}
\usepackage{amssymb}
\usepackage{mathtools}
\usepackage{fancyhdr}
\usepackage{multicol}
\usepackage{multirow}

\usepackage{mdframed}% http://ctan.org/pkg/mdframed
\usepackage[usenames,dvipsnames,svgnames,table]{xcolor}% http://ctan.org/pkg/xcolor
\usepackage{array}
\usepackage[%
  colorlinks=true,%
  linkcolor=black,%
  urlcolor=black,%
  citecolor=black%
]{hyperref}
\usepackage{nameref} % to enable reference to sec/chap by name
\usepackage{ifthen} % include logic
\usepackage[font={small,sl}]{caption}
\usepackage[authoryear]{natbib}
\renewcommand\bibname{References}
\newcommand{\mychapbib}{% use this at end of each chapter
  \addcontentsline{toc}{section}{\bibname}
  \bibliographystyle{natbib}
  \bibliography{strucbioinf}
}
\usepackage[colorinlistoftodos]{todonotes}
\usepackage{letltxmacro}

\usepackage{tocstyle} % to set toc raggedright
\usetocstyle{standard}
\settocfeature{raggedhook}{\raggedright}
\newcommand{\bibfont}{\small\raggedright}
\def\cite{\citep}

\LetLtxMacro{\oldTodo}{\todo}
\renewcommand{\todo}[2][]{\oldTodo[#1]{TODO: #2}}

\usepackage[normalem]{ulem}
\newcommand\inwish[1]{\oldTodo[inline,color=SkyBlue]{WISH: #1}}

\newboolean{onechapter}

\newcommand{\AF}[1][~]{K.\@#1Anton#1Feenstra}
\newcommand{\SA}[1][~]{Sanne#1Abeln}

\newcommand{\AJ}[1][~]{Annika#1Jacobsen}
\newcommand{\HM}[1][~]{Halima#1Mouhib}

\newcommand{\JvG}[1][~]{Juami#1H.\@#1M.\@#1van#1Gils}

\newcommand{\AM}[1][~]{Ali#1May}

\newcommand{\EvD}[1][~]{Erik#1van#1Dijk}

\newcommand{\JB}[1][~]{\mbox{Jochem}#1\mbox{Bijlard}}

\newcommand{\RH}[1][~]{\mbox{Reza}#1\mbox{Haydarlou}}

\newcommand{\IH}[1][~]{\mbox{Isabel}#1\mbox{Houtkamp}}

\newcommand{\orcid}[1]{\href{https://orcid.org/#1}{\raisebox{-0.7ex}{\protect\includegraphics[height=3ex]{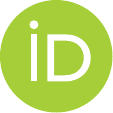}}}}
\definecolor{idgreen}{RGB}{166 206 57}
\newcommand{\mailid}[1]{\href{mailto:#1}{\raisebox{-0.3ex}{\color{idgreen}\textsf{\textbf{\Large \protect@}}}}}

\newcommand{\AFid}{\orcid{0000-0001-6755-9667}}
\newcommand{\SAid}{\orcid{0000-0002-2779-7174}}
\newcommand{\HMid}{\orcid{0000-0001-5031-3468}}

\newcommand{\JvGid}{\orcid{0000-0003-3706-7818}}

\newcommand{\RHid}{\orcid{0000-0003-4138-7179}}

\newcommand{\AJid}{\orcid{0000-0003-4818-2360}}
\newcommand{\EvDid}{\orcid{0000-0002-6272-2039}}

\newcommand{\AMid}{\orcid{0000-0002-0551-9966}}

\newcommand{\IHid}{\orcid{0000-0002-4222-7292}}

\newcommand{\JBid}{\mailid{j.bijlard@gmail.com}}

\newcommand{\ACtxt}{Wrote the text}
\newcommand{\ACfig}{Created figures}
\newcommand{\ACref}{Review of current literature}
\newcommand{\ACeds}{Editorial responsibility}
\newcommand{\ACproof}{Critical proofreading}
\newcommand{\ACfb}{Non-expert feedback}

\newcommand{\Angs}[1][~]{\text{\normalfont\AA}}

\renewcommand{\and}{\quad}

\newcommand{\pdbref}[1]{\href{http://www.rcsb.org/pdb/explore.do?structureId=#1}{PDB:#1}}
\newcommand{\arxiv}[2][UNDEFINED]{\href{https://arxiv.org/abs/#2}{\ifthenelse{\equal{#1}{UNDEFINED}}{arxiv.org/abs/#2}{#1}}}

\newcommand{\figref}[2][]{\hyperref[fig:#2]{Figure\@~\ref*{fig:#2}#1}}
\newcommand{\tabref}[1]{\hyperref[tab:#1]{Table \ref*{tab:#1}}}
\renewcommand{\eqref}[2][]{\hyperref[eq:#2]{Equation#1\@~\ref*{eq:#2}}}
\newcommand{\panelref}[2][]{%
    \ifthenelse{\boolean{onechapter}}{%
        \hyperref[panel:#2]{Panel\@~``\nameref{panel:#2}#1''}%
    }{%
        \hyperref[panel:#2]{Panel\@~\ref*{panel:#2}#1}%
    }%
}
\newcommand{\secref}[2][n]{%
    \hyperref[sec:#2]{%
        \ifthenelse{\equal{#1}{n} }{Section\@~\ref*{sec:#2}}{}% just number
        \ifthenelse{\equal{#1}{nn}}{Section\@~\ref*{sec:#2} ``\nameref{sec:#2}''}{}% nm & nr
        \ifthenelse{\equal{#1}{N} }{``\nameref{sec:#2}''}{}% just quoted name
        \ifthenelse{\equal{#1}{NN} }{\nameref{sec:#2}}{}% just name
    }%
}
\newcommand{\chref}[2][n]{%
    \ifthenelse{\boolean{onechapter}}{%
        \ifthenelse{\equal{#2}{ChPref}     }{\arxiv[Chapter ``\nameref*{ch:#2}'']{1801.09442}}{}%
        \ifthenelse{\equal{#2}{ChIntroPS}  }{\arxiv[Chapter ``\nameref*{ch:#2}'']{1801.09442}}{}%
        \ifthenelse{\equal{#2}{ChDetVal}   }{\arxiv[Chapter ``\nameref*{ch:#2}'']{2108.02706}}{}%
        \ifthenelse{\equal{#2}{ChStrucAli} }{\arxiv[Chapter ``\nameref*{ch:#2}'']{1801.09442}}{}%
        \ifthenelse{\equal{#2}{ChDBClass}  }{\arxiv[Chapter ``\nameref*{ch:#2}'']{1801.09442}}{}%
        \ifthenelse{\equal{#2}{ChFunc}     }{\arxiv[Chapter ``\nameref*{ch:#2}'']{1801.09442}}{}%
        \ifthenelse{\equal{#2}{ChIntroPred}}{\arxiv[Chapter ``\nameref*{ch:#2}'']{1712.00407}}{}%
        \ifthenelse{\equal{#2}{ChHomMod}   }{\arxiv[Chapter ``\nameref*{ch:#2}'']{1712.00425}}{}%
        \ifthenelse{\equal{#2}{ChSSPred}   }{\arxiv[Chapter ``\nameref*{ch:#2}'']{1801.09442}}{}%
        \ifthenelse{\equal{#2}{ChFuncPred} }{\arxiv[Chapter ``\nameref*{ch:#2}'']{1801.09442}}{}%
        \ifthenelse{\equal{#2}{ChIntroDyn} }{\arxiv[Chapter ``\nameref*{ch:#2}'']{1801.09442}}{}%
        \ifthenelse{\equal{#2}{ChThermo}   }{\arxiv[Chapter ``\nameref*{ch:#2}'']{1801.09442}}{}%
        \ifthenelse{\equal{#2}{ChMD}       }{\arxiv[Chapter ``\nameref*{ch:#2}'']{1801.09442}}{}%
        \ifthenelse{\equal{#2}{ChMC}       }{\arxiv[Chapter ``\nameref*{ch:#2}'']{1801.09442}}{}%
    }{% else
    \hyperref[ch:#2]{%
        \ifthenelse{\equal{#1}{n} }{Chapter \ref*{ch:#2}}{}% just number
        \ifthenelse{\equal{#1}{nn}}{Chapter \ref*{ch:#2} ``\nameref{ch:#2}''}{}% name & number
        \ifthenelse{\equal{#1}{N} }{``\nameref{ch:#2}''}{}% just name
      }%
  }%
}
\newcommand{\chrefname}[1]{\hyperref[ch:#1]{Chapter \ref*{ch:#1} ``\nameref{ch:#1}''}}
\newcommand{\partref}[1]{\hyperref[#1]{Part \ref*{#1}}}
\newcommand{\appref}[1]{\hyperref[app:#1]{Appendix \ref*{app:#1}}}

\newcommand{\figsource}[1]{\protect\footnote{Figure source location: \url{#1}}}

\newenvironment{cenum}[1][\itshape i)\upshape\ ]
{\begin{compactenum}[#1]} {\end{compactenum}}

\renewcommand{\arraystretch}{1.3}

\makeatletter \@beginparpenalty=5000 \makeatother

\newenvironment{bgreading}[1][]{
  \begin{mdframed}[%
      outerlinewidth=0,%
      linecolor=CornflowerBlue!30,%
      backgroundcolor=CornflowerBlue!30,%
      innerleftmargin=14,%
      innerrightmargin=14,%
    ]
	\ifthenelse{\equal{#1}{}}{}{% only if optional arg not empty
        \stepcounter{panel}
    	\subsection*{#1} % without panel numbers
    }
}{%
  \end{mdframed}
}

\usepackage{listings}
\definecolor{backcolour}{rgb}{0.95,0.95,0.92}
\definecolor{codegreen}{rgb}{0,0.6,0}
\definecolor{codegray}{rgb}{0.5,0.5,0.5}
\definecolor{codered}{rgb}{0.8,0,0.0}
\definecolor{codeblue}{rgb}{0.0,0,0.8}
\lstdefinestyle{codeStyle}{
    backgroundcolor=\color{backcolour},   
    commentstyle=\color{codegreen},
    keywordstyle=\color{codeblue},
    numberstyle=\tiny\color{codegray},
    stringstyle=\color{codegray},
    numbers=left,                    
    tabsize=2
} 
\makeindex

\begin{document}

\setboolean{onechapter}{true}

% page head and foots (feet):
\pagestyle{fancy}
\lhead[\small\thepage]{\small\sf\nouppercase\rightmark}
\rhead[\small\sf\nouppercase\leftmark]{\small\thepage}
\newcommand{\innerfoot}{\footnotesize{\sf{\copyright} Feenstra \& Abeln}, 2014-2023}
\newcommand{\outerfoot}{\footnotesize \sf Intro Prot Struc Bioinf}
\lfoot[\outerfoot]{\innerfoot}
\cfoot{}
\rfoot[\innerfoot]{\outerfoot}
\renewcommand{\footrulewidth}{\headrulewidth}

\mainmatter
\setcounter{chapter}{11}
\chapterauthor{\JvG*~\JvGid \and \EvD~\EvDid  \and \AM~\AMid \and \HM~\HMid \and \JB~\JBid \and \AJ~\AJid \and \IH~\IHid \and \AF*~\AFid \and \SA*~\SAid }
\chapterfigure{\includegraphics[width=0.5\linewidth]{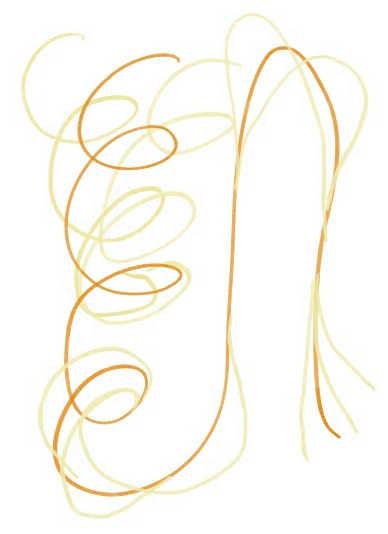}}
\chapterfootnote{* editorial responsability}
\chapter{Introduction to Protein Folding}
\label{ch:ChIntroDyn}

\ifthenelse{\boolean{onechapter}}{\tableofcontents\newpage}{} 

In this chapter we explore basic physical and chemical concepts required to understand protein folding. We introduce major (de)stabilising factors of folded protein structures such as the hydrophobic effect and backbone entropy.  In addition, we consider different states along the folding pathway, as well as natively disordered proteins and aggregated protein states. In this chapter, an intuitive understanding is provided about the protein folding process, to prepare for the next chapter on the thermodynamics of protein folding. In particular, it is emphasized that protein folding is a stochastic process and that proteins unfold and refold in a dynamic equilibrium. The effect of temperature on the stability of the folded and unfolded states is also explained.

\section{Protein folding and restructuring}
\subsection{Flexibility of protein chains \& structural ensembles}

\begin{figure}[b]
\centerline{
  (a)
  \includegraphics[width=0.5\linewidth]{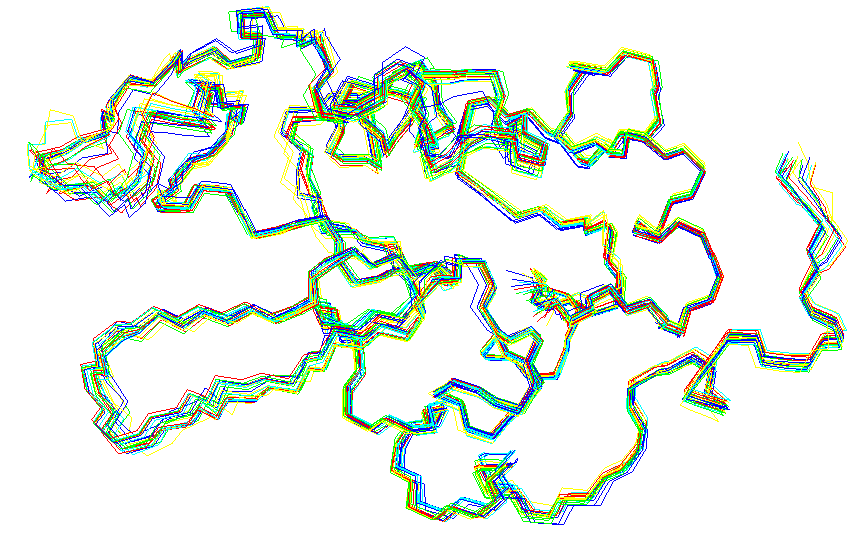}
  (b)
  \includegraphics[width=0.5\linewidth]{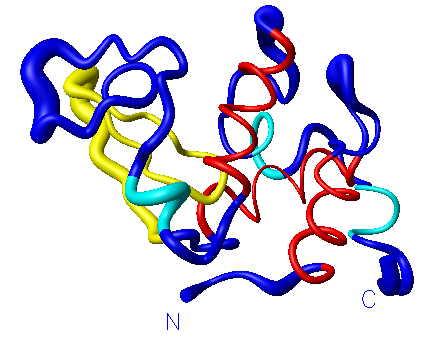}
}
\caption{Proteins do not necessarily take one single structural conformation, but may instead be flexible. The native, functional state may contain many different structural conformations. (a) An ensemble of conformations based on NMR experimental data, shown as a set of overlaid backbone traces. (b) Another ensemble of conformations where the variation is shown as the thickness of the backbone - this is also known as the sausage representation.}
\label{fig:ChIntroDyn-ensemble}
\end{figure}

In structural biology, it is generally believed that the protein sequence determines the 3D structure, which determines the function of the protein. Thus, a protein acquires its function once it is folded into its three-dimensional structure, the native state. This provides us with a very rigid view of protein structures. As we have seen in the previous chapters of this book, many methods in structural bioinformatics rely on this rigid view; examples are structure prediction, structure comparison and structure validation. However, proteins should in fact be viewed as flexible molecules that can take up a whole ensemble of different structural conformations (see \figref{ChIntroDyn-ensemble} for a simple illustration of a conformational ensemble for a protein structure). This ensemble may change depending on physical conditions such as temperature, pH, salt concentration, or the concentration of the protein in question. The conformational ensemble also may change depending on the presence of binding partners (e.g. small ligands, DNA or other proteins), a membrane, or crowding of the cell cytoplasm. In fact, exactly this dynamic interaction between the structural ensemble of a protein with its environment, is what allows for the functionality of proteins. Several examples of this are discussed in some detail in \chref[nn]{ChFunc}. 

It is not difficult to see why it is important to take the flexibility of proteins into account; for example protein folding, allostery (conformational change upon ligand-binding) and complex formation would not be possible without a protein changing its shape. However, most experimental and bioinformatics methods described in the previous chapters cannot (explicitly) deal with this flexibility. In this part of the book, we will study \emph{simulation methods}, and explain the underlying \emph{thermodynamic principles} behind these simulations. Moreover, we will consider how protein flexibility can be modeled explicitly: simulations allow us to consider the ensemble of different structural conformations of a protein. Simulations, in combination with experimental observations, can give us insight into the process of folding and, perhaps more importantly, investigate how the flexibility of its structure allows a protein to perform its function. We should therefore not think of protein structure and folding as deterministic or static phenomena, but as stochastic and dynamical processes. 

\subsection{Defining the folded and unfolded states}
Before we go any further, it may be helpful to define the folded, or \emph{native state} of a protein. Intuitively, you may have a good idea what such a folded state looks like, as most of the experimentally determined structures resemble a uniquely folded conformation (if the experiment was performed at very low temperatures); the folded state actually covers a small ensemble of conformations at physiological temperatures. In fact, it is the unfolded state that may be less intuitive: this state covers a large ensemble of (possibly extended) conformations of the peptide backbone. The exact nature of this ensemble may depend on the specific conditions in the system. For example, at low temperatures, the unfolded state may be more compact than at high temperatures \cite{vanDijk2016}.

There are different experimental methods that can observe if a solution contains folded proteins, e.g. NMR (Nuclear Magnetic Resonance) or CD (Circular Dichroism) \cite{Wuthrich19896Overview, Kelly2005HowDichroism}. For example, NMR can be performed in solution, allowing full structural information to be resolved for small proteins; but only if the conformations are similar (hence when a protein is folded).  Alternatively, more indirect measurements can determine if the proteins in a solution are fully folded: for example, Green Fluorescent protein (GFP) will only show fluorescence when fully folded. This protein is therefore often used for folding experiments. Similarly, enzymes typically only show catalytic activity in their fully folded native form; for enzymes, enzymatic activity can report if the protein is folded.

In simulations on the other hand, we typically compare a simulation snapshot, or conformation, to the experimentally determined structure in order to see if it is folded. Two commonly used measures to indicate if a protein is folded are: 1) RMSD to the native structure - determined by superpositioning the conformation onto the native structure (see \chref{ChStrucAli}) or 2) the topological similarity of internal contacts - calculated by taking the intersection between the contact map of the conformation and the contact map of the native structure. 

By analysing these measures over a simulation run, we can see that the number of possible conformations that is similar to a particular native structure is much lower than the number of conformations that are dissimilar to the native structure, with the latter ensemble of conformations defining the unfolded structure.

\section{Folding and refolding}

\begin{figure}
\centerline{
  \includegraphics[width=0.9\linewidth]{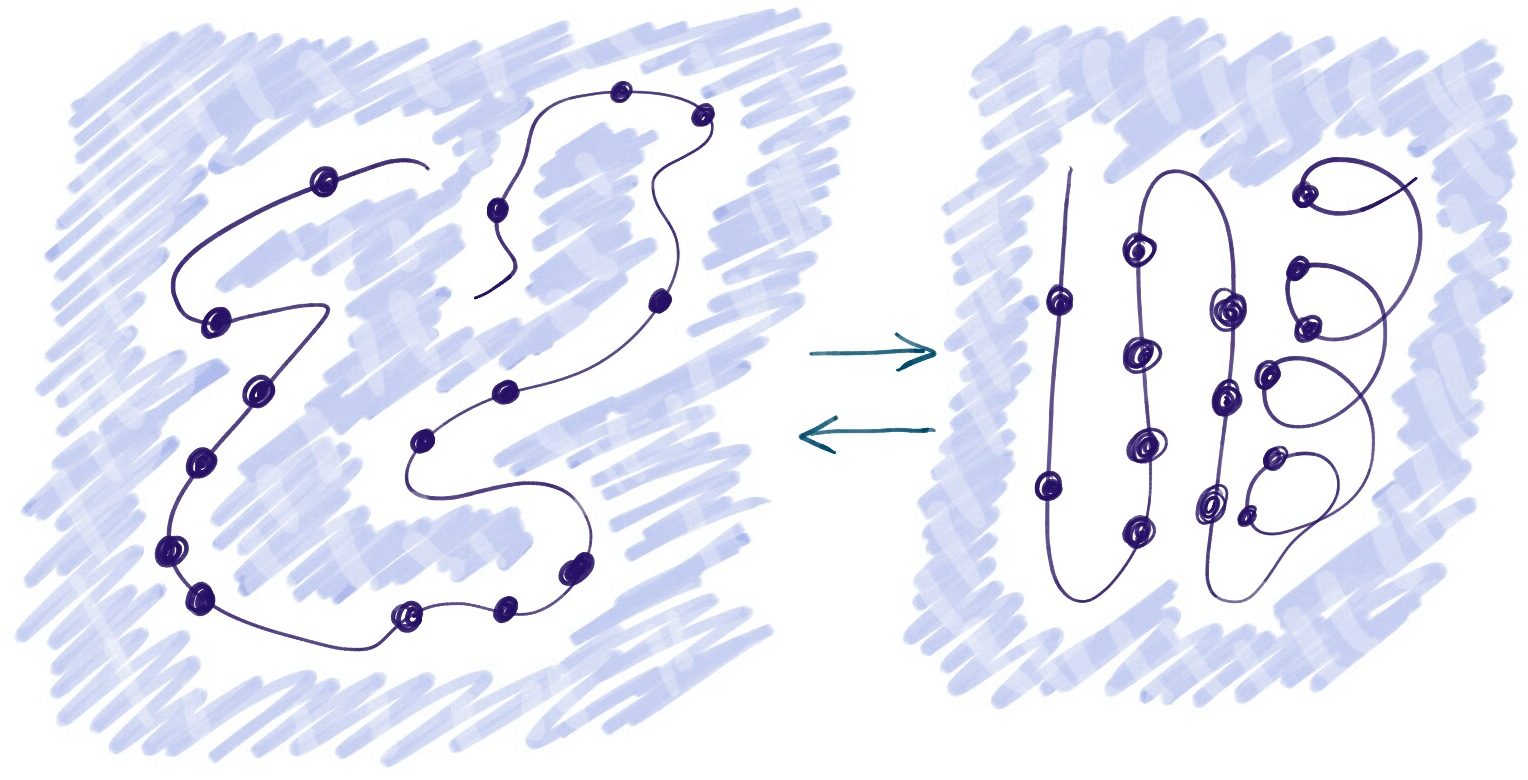}
}
\caption{Denatured, unfolded protein chain on the left and the native, folded state on the right. The protein is shown in dark blue, water is shown in light blue. Dots on the protein indicate hydrophobic residues. One may observe that in the unfolded state interactions with the water (solvent) are far more extensive; more precisely there is a large interface between the solvent and the residues in the protein. In the folded state, only the outside of the protein interacts with the water, while hardly any solvent is present in the core of the protein. This is a result of the hydrophobic effect (see \secref{ChIntroDyn:hydrophobiceffect}).}
\label{fig:ChIntroDyn:Fold}
\end{figure}

To further illustrate the dynamic and flexible nature of protein molecules, we will first try to sketch a picture of what we mean by protein folding.  %As this is conceptually not straightforward, we will first consider a somewhat simplified case. 
Consider we have a simple, \textsl{in vitro}, system of proteins dissolved in water (see \figref{ChIntroDyn:Fold}). These proteins, all with the same sequence, are two-state folders, meaning they are only stable in the folded or unfolded state. The folding and unfolding rate of the proteins is rapid, such that the proteins will reversibly unfold and refold over time.
Moreover, the system is in a dynamic equilibrium, meaning that the fraction of unfolded proteins (and therefore also the fraction of folded proteins) remains stable over time (see the \chref{ChThermo} for a detailed explanation of equilibrium). On a path from unfolding to folding, a single molecule will visit a large number of different conformations. Nevertheless, it will spend the majority of its time either in the folded or unfolded state.

\subsection{Stability and probability}
\begin{figure}
\centerline{
  \includegraphics[width=0.9\linewidth]{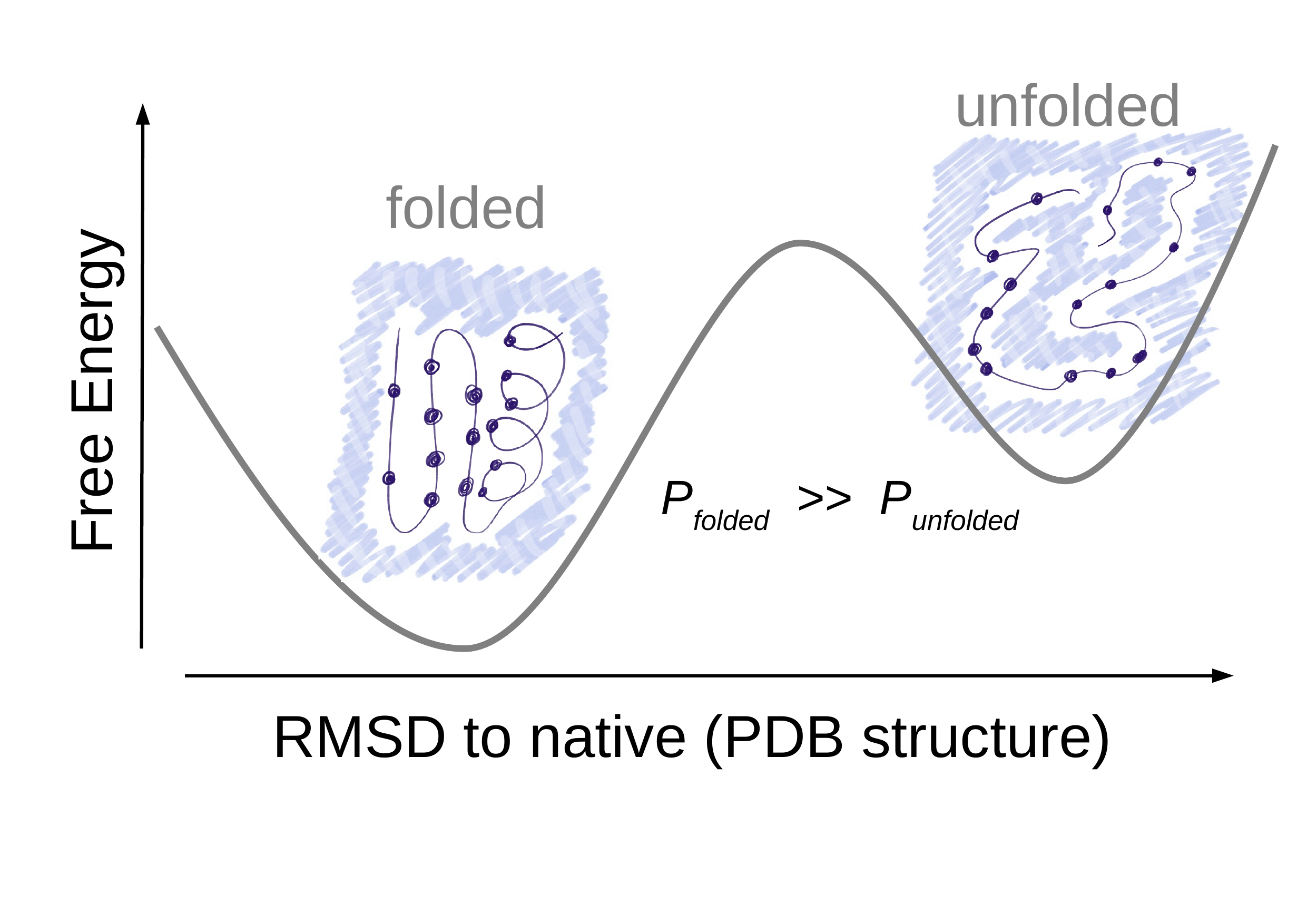}
}
\caption{Sketch of a free energy landscape for a protein under physiological conditions. The protein is said to spend most of its time in the native or folded state (left well, low RMSD to native), as this state has the lowest free energy. Note that under these conditions, the native state is not exactly the same as the PDB structure but nevertheless very similar. The other local minimum (right well, high RMSD to native) represents the unfolded state. $P_\text{folded}$ is the probability to find the protein in the folded state, which here is higher than $P_\text{unfolded}$: the the probability to find the protein in the unfolded state.}
\label{fig:ChIntroDyn-FreeEnergy}
\end{figure}

Now, one of the most important observations to make is the relation between the \emph{probability} of being in a certain state and the \emph{stability} of that state in our simple system. Typically, under physiological conditions, around 30 $^{\circ}$C, proteins would be most stable in their native, folded state. This means that the chance of finding a single molecule in the folded state would be much higher than finding it in the unfolded state; see \figref{ChIntroDyn-FreeEnergy}. Moreover, the fraction of folded molecules will, under these conditions, be much higher than the fraction of unfolded proteins. 
Another way of phrasing this is that the \emph{free energy} of the folded state is lower that the free energy of the unfolded state. In \chref{ChThermo} we will see that the probability of finding a molecule in a given state, is quantitatively directly related to the free energy of that state.

\subsection{Changing conditions}
In the system described above we can change physical conditions, for example the temperature. If we were to raise the temperature to about 70 $^{\circ}$C, the unfolded state would become more stable than the folded state. In the next chapter we will see that this is an \emph{entropic} effect of the peptide backbone: at higher temperatures states that are comprised of many possible conformations (in this case the unfolded state) are favoured. Note that other changes in conditions, such as altering the pH, salt concentration or adding a denaturant, e.g.\@ urea, may also have a strong effect on the relative stabilities of the folded and unfolded states \cite{McNay2001ProteinStrength, Sahin2010ComparativeAntibodies}.

Finally, please note that in an equilibrium situation the fractions of folded and unfolded conformations will be completely determined by the probability of these states, under the given conditions. However, when equilibrium has not been reached we get an unstable situation: for example, if a protein solution has just been heated up very quickly  (known as a temperature jump experiment)\cite{French19691Method},  the fraction of unfolded protein molecules is still small, even though the unfolded state may be more favourable under these new conditions. In this case, the system will relax over time, until an equilibrium is reached again.

\section{Factors that (de)stabilize the native fold}
So, why and how do proteins fold into their unique native structures? What are the important physical factors contributing to the transition of an unstructured polypeptide chain to a specific three-dimensional shape? 

Since we know that the folded state should be the most stable state under native conditions, we can rephrase this question: why is the native state more stable than the unfolded state? There are many factors that contribute to the stability of the native state. 

\subsection{Hydrophobic effect}
\label{sec:ChIntroDyn:hydrophobiceffect}
It is thought that for the majority of globular proteins, the burial of hydrophobic side chains is the most important stabilizing effect on the protein structure \cite{Tsong1972,Baldwin2007}: water molecules are strongly attracted to each other, due to the possibility to form intermolecular hydrogen bonds between the electropositive H and electronegative O atoms. Hydrophobic particles, cannot form such hydrogen bonds. Therefore a water molecule generally prefers to be situated among other water molecules, instead of forming in interface with a hydrophobic substance. In the unfolded state, many hydrophobic residues will form an interface with the water (see the left panel of \figref{ChIntroDyn:Fold}). In the folded state, on the other hand, the hydrophobic side chains do not form an interface  with the water (see the right panel of \figref{ChIntroDyn:Fold}). Hence, with respect to the hydrophobic effect, the folded state is most favourable.

\subsection{Hydrogen bonds, salt-bridges and packing}
Other important factors that may stabilize the native state are van der Waals forces between the side chain atoms, hydrogen bonds between side chain atoms and/or backbone atoms, and salt bridges between charged side chains \cite{Baldwin2007}. Lastly, also some quantum effects can stabilize protein structures, a good example are the $\pi$-$\pi$ interactions formed by aromatic residues. \chref[nn]{ChIntroPS} explained many of these interactions in detail, and in \chref[nn]{ChMD} we will describe some more detail on how these interactions can be modeled and calculated.

\subsection{Backbone entropy}
Lastly, there is also a factor that favours the unfolded state: the entropy of the backbone. In the next Chapter we consider this effect in more detail. For now, it is enough to understand that there are way more possibilities to generate an unfolded conformation than to exactly match the native conformation of a protein. So if none of the factors (e.g.\@ hydrophobic effect and backbone hydrogen bonding) mentioned in the previous sections would favour the folded state, all proteins would be unfolded.

\begin{bgreading}[Anfinsen's Theorem]
Environmental factors such as solvent properties, temperature and pH are known to contribute to the specific three-dimensional structures of the native protein. However, the most important determinant of the folded structure is the amino acid sequence. Anfinsen showed, in his Nobel Prize winning denaturation-renaturation experiments, that proteins can be denatured and then will spontaneously refold to their native forms when conditions are changed back \cite{Anfinsen1973}. These findings resulted in the general acceptance of what is now called the ``thermodynamic hypothesis'', which states that the folded structure of a protein is fully encoded by its sequence, and the protein finds this structure due to thermodynamic laws. 

We can rephrase the idea behind this theorem by saying that the folded state is the most likely, lowest free energy, state in the native conditions.  For this to hold there are three important conditions: 
\begin{cenum}
\item uniqueness of the free energy minimum, given the sequence,
\item stability of the free energy minimum, 
\item kinetic accessibility of the free energy minimum. 
\end{cenum}
Point i) suggests that a (naturally evolved) protein sequence, folds specifically into a specific structure; in other words the sequence is the recipe for the exact structure the protein will take. Point (iii) suggests that the folding and unfolding rates are sufficiently high or - in other words - that the barrier between the unfolded and the native state should not be too high. High barriers may in practice prevent a protein from reaching the folded state.
Note that some proteins may require special conditions to fold; we will return to this in \secref{ChIntroDyn:seccell}.
In the last section of this Chapter we will see that, although the thermodynamic hypothesis seems to hold true for most naturally evolved proteins, there are also many proteins where the functional state, which is observed in nature, is not the one with the lowest free energy measured in \textsl{in vitro} in experiments. 
\end{bgreading}

\section{Folding pathways}
Previously, we considered a the case of a two-state folder: a protein for which the folding pathway only contains two stable states: the folded and unfolded state; moreover we assumed fast transitions between the folded and unfolded state. For many proteins, the folding pathways may be more complex, with intermediate stable states. For example, a multi-domain protein may fold one domain at a time, in a specific order. The state, in which only the first domain is folded would typically be a (meta-)stable state; this state is extremely likely to be visited on the path from the unfolded to the folded state and vice versa. More recently, it has been shown that for several proteins there exist smaller intermediate folding structures, or foldons, that appear as meta-stable states on the folding path \cite{Englander2017}. 

It is important to note that for different proteins, very different pathways have been observed experimentally \cite{Hartl2009,Dobson2003}. Moreover, generally folding and refolding is a stochastic process \cite[e.g.][]{Baclayon2016}. We will look at this in more detail in the last section of this Chapter.

\subsection{Free Energy Landscapes}

We have already shown a simple free energy landscape \figref{ChIntroDyn-FreeEnergy}. Many processes, like protein folding or protein-protein interactions, can be described as two-state processes. That means, there are two free energy minima, which are separated by a barrier. The top of the barrier is referred to as the \emph{transition state}, i.e.\@ the state through which the system must progress to go from one state to another. The height of the barrier determines the rate of the transition from one state to another. This makes knowing the barrier height important. However, from simulations it typically is difficult to sample a barrier, because the transition state is often very unstable and therefore rarely visited. The system will spend most of its time in the lowest of the two free energy minima (this is true for simulations and experiments). We will go in detail into the relation between free energy and probabilities in the next chapter, \chref[N]{ChThermo}. Even though a folded protein will also visit the other (unfolded) minimum, states that cross the barrier are short-lived. There are several techniques to improve the sampling of the transition state in simulations, which we will return to in \chref[nn]{ChMC}.

\begin{bgreading}[Levinthal's paradox]
How does this spontaneous folding occur? Levinthal argued that if a small protein would have to sample every possible three-dimensional conformation before obtaining its native structure, it would take more time than the age of the universe for it to find its native structure \cite{Levinthal1969}. To understand this, let us consider a protein of 100 amino acids, where each peptide bond in between two amino acids has two possible torsion angles, and each of these angles can assume three different values. The protein then has $3^{99 \times 2} \approx 2.9 \times 10^{94}$ possible conformations. If each conformation can be visited in one picosecond ($10^{-12}s$) it would take about $10^{75}$ years for the protein to to visit all possible conformations (our universe is $13.8 \times 10^9$ years old). 

However, it is known that small proteins like this can fold into their native structures in a matter of seconds. This phenomenon, known as ``Levinthal's paradox'', suggests that the folding protein only samples a very small fraction of all possible conformations before it finds its most stable state. %Instead, you may think the protein as being `guided' along some path from the unfolded to the folded state. 
A typical protein would have a path towards the folded state that is relatively smooth, without very large barriers. Such a folding path may allow early formation of stable interactions, allowing the molecule to obtain its lowest energy state within reasonable time. Note that inside the cell, other factors, such as chaperones, may in fact make a folding path more smooth, see section \secref{ChIntroDyn:seccell}.
\end{bgreading}

\section{Folding in the cell}
\label{sec:ChIntroDyn:seccell}
Inside the cell (\textsl{in vivo}), the folding process occurs during and after the synthesis of the polypeptide chains in the ribosomes. The correct folding is necessary for the protein to perform its biological function. % \figref{ChIntroDyn:Fold}.
Up until this point, we have assumed protein folding to be a reversible process. Note that in practice some proteins are not able to refold \textsl{in vivo} after unfolding (or denaturation); some proteins only fold directly while being synthesized at the ribosome, and some proteins require chaperones to fold from an unfolded state. Other proteins, or protein regions, may only fold, upon binding a specific binding partner.

\subsection{Chaperones}

To prevent misfolding, folding \textsl{in vivo} is often aided by chaperones. GroEL is one of them, but there are several others \cite{Horwich2006GroEL-GroES-mediatedFolding}. Most chaperones, including GroEL, are so-called heat-shock proteins, which were given this name because bacteria upregulate them as temperature increases. This makes sense as the chances of protein misfolding and aggregation increases with temperature. For many proteins this aid from chaperones is a necessity for reaching the native state or for refolding after denaturation. Chaperones have several functions in the cell. In addition to folding, there are also chaperones that prevent aggregation of misfolded proteins.

\subsection{Folded proteins are only marginally stable}
Most naturally occurring proteins are only marginally stable under physiological conditions \cite{Privalov1974, Pucci2017}. This means that there is only a small free energy difference between the folded and unfolded state. This is most likely the result of evolution: proteins only need to be stable enough to perform their function, and have thus found a balance between stability and flexibility in order be able to move and function. In fact, making them more stable may result in an additional cost when proteins need to be `cleaned up' by degradation in the ubiquitin-mediated proteasome \cite{Wilson2020EvolutionaryStable}.

\section{Alternative stable states of proteins}

Besides the folded and unfolded state, there may be alternative (meta)stable states for a protein. As previously suggested, some of these may lie on the pathway from the folded to the unfolded state. There are also some states that, depending on the conditions, may actually compete with the native functional state. Some important alternative states are listed below.

\subsection{Molten globules}
\begin{figure}
\centerline{\includegraphics[width=1.2\linewidth]{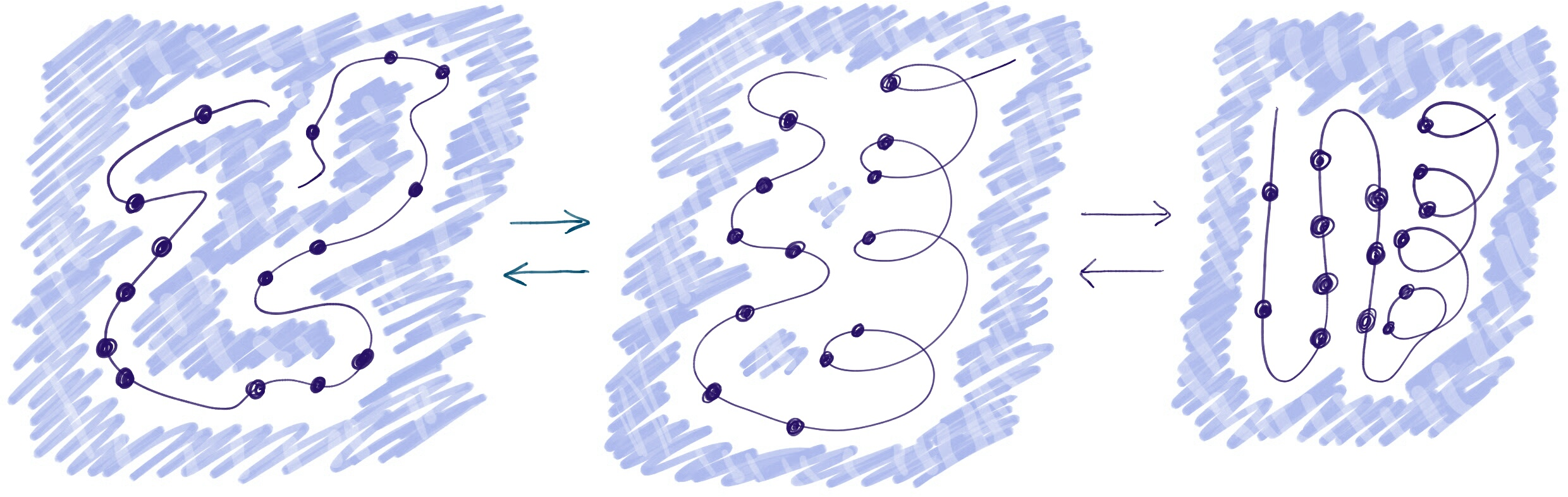}}
\caption{The transition from denatured (on the left) to folded (on the right) goes via some transition state which involves a hydrophobic collapse: all (or most) of the hydrophobic residues (here drawn as circles) are on the inside, but not all of the secondary structure has been formed yet. This intermediate state is often referred to as `molten globule'.}
\label{fig:ChIntroDyn-hydrophobic-collapse}
\end{figure}
So far we have considered a single unfolded state. In fact, different types of unfolded states can be defined ranging from states with mostly extended conformation, to states that are considerably compact. If such states are separated by a free energy barrier - we can truly observe different properties both by experiments and simulations.

The molten globule state is a compact state in which hydrophobic amino acid residues are clustered as drawn schematically in \figref{ChIntroDyn-hydrophobic-collapse}. Local hydrophobic groups are formed to avoid unfavorable interactions with water. However, the core of the protein may still be somewhat more permeable than the fully folded state, because this state is typically less compact than the fully folded state and not all secondary structure elements are formed \cite{Baldwin2013}. Molten globule states may be observed under specific conditions, in which the molten globule state is more stable than the folded state. Alternatively molten globules may be  meta-stable states along a folding pathway \cite{Hartl2009,Dijkstra2018}. This latter concept is also referred to as `the hydrophobic collapse'.

\subsection{Natively disordered proteins}
Proteins for which the functional conformation is (largely) unfolded are called natively, or intrinsically, disordered proteins. Such proteins were already discussed in \chref[nn]{ChFunc} \secref{ChFunc:disorderedProteins}.

\subsection{Misfolding}
\textit{In vivo}, cellular processes may occasionally fail, including protein folding. Some proteins may end up in a misfolded state. Such a state would not be functional, but may be rather stable. From these states the barrier to a correctly folded state may be so high, that a transition is very unlikely, in this scenario the protein is kinetically trapped in a misfolded state. 

Proteins may become misfolded by chance,  through a `casualty' on the folding pathway, through an abrupt change in conditions (e.g.\@ heat shock) , or through interaction with other proteins (e.g.\@ prions).

Misfolded proteins could be very dangerous in the cell, as many hydrophobic residues may be exposed to the surface. In a crowded environment, such a misfolded protein would be very sticky, and could disrupt other parts of the cell (e.g.\@ by making a membrane leaky or sticking to other proteins) \cite{Relini2014ProbingBilayers, Bondarev2018ProteinClassification}.

\subsection{Aggregation and amyloid formation}
An even more dangerous form of misfolding, is where several protein molecules start aggregating together. 
Specific forms of such aggregates are amyloid fibrils, where multiple chains of proteins form large beta sheets \cite{Chiti2006,Dobson2003}. 
This amyloid state, is - under several conditions - actually more stable than the folded state \cite{Buell2014}. Meaning that if you wait long enough (think years), many proteins would end-up in amyloid fibrils.

In \chref{ChIntroPS}, %\panelref{ChIntroPS:atypical-ss} 
we gave an example of misfolding of prion proteins, which leads to formation of $\beta$ fibrils that disrupt cellular function and even kills cells. In general, misfolding of proteins is a problem that cells need to avoid. Therefore, in order to protect other elements in cell from misfolded proteins, a cell typically has an extensive machinery to aid folding, or to target misfolded and/or aggregated proteins for degradation. Chaperones (discussed in the previous section), are a part of this machinery.

\begin{bgreading}[Protein folding in experiment and simulation]

It is not straightforward to study the folding of proteins, neither in experiment nor in simulation. Specifically, it is extremely difficult to observe intermediate stages of the folding pathway. These intermediates will typically not be very stable and therefore are only present for short times, and at low concentrations. Experimental procedures, such as X-ray crystallography, NMR and various spectroscopic methods typically give information on structure, but limited detail on dynamical processes like folding (see also \chref[nn]{ChDetVal} for an overview of these methods). Nevertheless, it is possible to obtain insight into which residues are important for intermediate states on the folding pathway. One trick is to see how the rate of folding changes, when mutations are made in a protein sequence. As the speed of folding is directly related to the height of the energy barrier between the folded and the unfolded states, one can infer if a mutation stabilizes or destabilizes the transition state (state on top of the folding barrier). This analysis, called `Phi ($\phi$) analysis' or `Alanine scanning', is therefore an important trick to get an insight into the transition state and folding nucleus \cite{Fersht}. You see an example of it in \citet{Shaw2010}, where it is applied to the small FiP35 $\beta$-sheet peptide folding.

On the other hand computational folding simulations offer a possibility to study the underlying mechanisms of protein folding. For example, molecular dynamics (MD) simulations may be used to study protein folding for some specific, small proteins \cite{Shaw2010}. However, generally speaking it is not possible to study the full folding of a protein using direct simulation techniques and fully detailed atomistic models. One of the reasons is that the length of the computational time needed to fold a protein is too large. Another problem is that the models used may not be accurate enough. To give an indication what is currently possible:  Molecular dynamics (MD) simulations were used to obtain milliseconds-scale folding events of a small peptide \cite{Daura1998} and small proteins \cite{Shaw2010}, however the folding of large proteins remains out of reach. More about molecular dynamics simulations and protein folding in \chref{ChMD}.

Nevertheless, the combination of experimental observations and computational simulation can give us very good insight into the nature of folding proteins \cite{Vendruscolo2005,Knowles2014,Fersht,Tompa2009,Shaw2010}.

\end{bgreading}

\section{Key concepts}
\begin{compactitem}
\item Proteins can take many different conformations
\item Under physiological conditions the native (functional) state is typically most stable
\item Proteins fold, unfold and re-fold continuously $\rightarrow$ dynamic equilibrium
\item Protein folding is a stochastic rather than a deterministic process
\item In the crowded environment of the cell, special precautions must be taken to allow proteins to fold properly, and to avoid problems due to accumulation of misfolded proteins.
\item Increasing the temperature increases the stability of the entropically favourable state. In protein folding, this is typically the unfolded state.
\item Decreasing the temperature increases the stability of the enthalpically favourable state. %In protein folding, this is typically the folded state (at least for temperatures significantly above 0 $^\circ$C).
\end{compactitem}

\section{Further reading}

\begin{compactitem}
\item ``Converging concepts of protein folding in vitro and in vivo'' -- \citet{Fersht}
\item ``Energetics of Protein Folding.'' -- \citet{Baldwin2007}
\item ``Physical and molecular bases of protein thermal stability and cold
adaptation.'' -- \citet{Pucci2017}
\item ``Protein folding and misfolding.'' -- \citet{Dobson2003} 
\item ``Protein misfolding, functional amyloid, and human disease.'' -- \citet{Chiti2006}
\end{compactitem}

\section*{Author contributions}
{\renewcommand{\arraystretch}{1}
\begin{tabular}{@{}ll}
\ACtxt: &   JvG, EvD, HM, KAF, AM, SA \\
\ACfig: &   JvG, JB, KAF, SA, \\
\ACref: &   JvG, IH, KAF, SA\\
\ACproof:&  AF, HM, JvG, SA \\
\ACfb:  &   JB, AJ \\
\ACeds: &  JvG, KAF, SA
\end{tabular}}

\noindent
The authors thank \RH~\RHid{} for non-expert feedback.

\mychapbib
\clearpage
%\addcontentsline{toc}{chapter}{\bibname}
%\bibliography{strucbioinf}

\cleardoublepage

\end{document}